\documentclass[journal=jacsat, manuscript=article]{achemso}
\pdfoutput=1
\usepackage[utf8]{inputenc}
\usepackage[english]{babel}
\usepackage{graphicx}
\usepackage{subfigure}
\usepackage{float}
\usepackage{amsmath}
\usepackage{amssymb}
\usepackage{bibentry}
\usepackage{amsfonts}
\usepackage{array}
\usepackage{listings}
\usepackage{lmodern}
\usepackage{mathpazo}
\usepackage{microtype}
\usepackage{hyperref}
\usepackage{xfrac}
\usepackage{color}

\usepackage{changes}

\hypersetup{
    colorlinks,%
   	citecolor=black,%
    filecolor=black,%
    linkcolor=black,%
    urlcolor=black}%
\usepackage[T1,OT1]{fontenc}

\author{Jung-Ching Liu}
\email{jungching.liu@unibas.ch}
\affiliation{Department of Physics, University of  Basel, Klingelbergstrasse 82, 4056 Basel, Switzerland}

\author{R\'emy Pawlak}
\affiliation{Department of Physics, University of  Basel, Klingelbergstrasse 82, 4056 Basel, Switzerland}

\author{Xing Wang}
\affiliation{Department of Chemistry, Biochemistry and Pharmaceutical Sciences, University of Bern, Freiestrasse 3, 3012 Bern, Switzerland}

\author{Philipp D'Astolfo}
\affiliation{Department of Physics, University of  Basel, Klingelbergstrasse 82, 4056 Basel, Switzerland}

\author{Carl Drechsel}
\affiliation{Department of Physics, University of  Basel, Klingelbergstrasse 82, 4056 Basel, Switzerland}

\author{Ping Zhou}
\affiliation{Department of Chemistry, Biochemistry and Pharmaceutical Sciences, University of Bern, Freiestrasse 3, 3012 Bern, Switzerland}

\author{Silvio Decurtins}
\affiliation{Department of Chemistry, Biochemistry and Pharmaceutical Sciences, University of Bern, Freiestrasse 3, 3012 Bern, Switzerland}

\author{Ulrich Aschauer}
\affiliation{Department of Chemistry, Biochemistry and Pharmaceutical Sciences, University of Bern, Freiestrasse 3, 3012 Bern, Switzerland}

\author{Shi-Xia Liu}
\affiliation{Department of Chemistry, Biochemistry and Pharmaceutical Sciences, University of Bern, Freiestrasse 3, 3012 Bern, Switzerland}

\author{Wulf Wulfhekel}
\affiliation{Physikalisches Institut, Karlsruhe Institute of Technology, Wolfgang-Gaede-Str. 1, 76131 Karlsruhe, Germany}

\author{Ernst Meyer}
\email{ernst.meyer@unibas.ch}
\affiliation{Department of Physics, University of  Basel, Klingelbergstrasse 82, 4056 Basel, Switzerland}

\date{\today}
\title{Proximity-Induced Superconductivity in Atomically Precise Nanographene}

\begin{document}
\begin{abstract}
\noindent 
Obtaining a robust superconducting state in atomically precise nanographene (NG) structures by proximity to a superconductor could foster the discovery of topological superconductivity in graphene. On-surface synthesis of such NGs has been achieved on noble metals or metal oxides, however, it is still absent on superconductors. Here, we present a synthetic method to induce superconductivity to polymeric chains and NGs adsorbed on the superconducting Nb(110) substrate covered by thin Ag films. Using atomic force microscopy at low-temperature, we characterize the chemical structure of each sub-product formed on the superconducting Ag layer. Scanning tunneling spectroscopy further allows us to elucidate electronic properties of these nanostructures, which consistently show a superconducting gap. We foresee our approach to become a promising platform for exploring the interplay between carbon magnetism and superconductivity at the fundamental level.\\


\end{abstract}

\section{Introduction}   

Topological superconductors are currently of particular interest in the condensed matter physics due to their potential as building blocks for topological quantum computation.~\cite{Stern2013,Sarma2015} Topological superconductivity (TS) is elusive in nature, but it can be engineered in hybrid heterostructures by coupling an electron gas with spin-momentum locking to a conventional superconductor.\cite{Fu2008,Sato2017,Lutchyn2018} As previously demonstrated in ferromagnetic atomic chains\cite{NadjPerge2014,Ruby2015,Pawlak2016} or islands\cite{Kezilebieke2020} proximitized with a $s$-wave superconductor, fingerprints of TS are Majorana quasiparticle excitations, so-called Majorana zero modes (MZMs). MZMs can be identified by scanning tunneling spectroscopy (STS) as zero-energy conductance peaks located at system boundaries which also mark the system topology.\cite{Jaeck2021}

With the bottom-up synthesis through on-surface reactions, atomically-precise nanogra-phenes (NGs) and graphene nanoribbons (GNRs)\cite{Cai2010,Ruffieux2016} can host Dirac fermions, topological electronic properties,\cite{Groning2018,Rizzo2018,Rizzo2020} magnetic edge states,\cite{Ruffieux2016,Mishra2020a} and coupled spins.~\cite{Li2019,Mishra2021} Despite considerable efforts to engineer their structures and electronic properties on surfaces, observing the interaction of a superconducting state with graphene local magnetism is scarce in literature.~\cite{Rio2021} Interestingly, this interaction can lead to further application from strong spin-orbit coupled materials to novel graphene-based topological superconductors~\cite{Hoegl2020}, which opens a new era for implementing Majorana-based $qubits$ in topological quantum computation.

The fabrication of NGs is based on a bottom-up approach in which predefined organic precursors are deposited under ultra-high vacuum (UHV) conditions onto crystalline substrates, and then undergo thermally triggered reactions. However, these reactions are so far restricted to noble metals (Au, Ag or Cu)~\cite{Cai2010,Simonov2018} or metal oxides (TiO$_2$),~\cite{Kolmer2020} where surface diffusion of molecules, dehalogenation and cyclodehydrogenation processes are possible by thermal treatment. In contrast, most conventional superconductors are unable to host these reactions due to their low melting points (Pb, In) or high reactivity (Nb, Re). Extending the on-surface chemistry toolbox to superconducting surfaces thus represents a crucial step in the study of the subtle interplay between carbon magnetism and superconductivity.

To date, proximity-induced superconductivity in epitaxial graphene has been achieved by top superconducting electrodes~\cite{Natterer2016}, by proximity to co-adsorbed Pb islands~\cite{Rio2021}, or by synthesizing graphene on a superconducting substrate such as Re(0001).\cite{Tonnoir2013} Although the last approach offers the cleanest alternative as it prevents the contamination of graphene during the fabrication process, the strong hybridization of graphene $\pi$ bands with $5d$ Re orbitals was found to prevent Dirac-like electronic dispersion close to the Fermi level.~\cite{Tonnoir2013} Later, this issue was circumvented by intercalating an Au buffer layer between graphene and the superconductor,\cite{Mazaleyrat2020} which restores the intrinsic electronic properties similar to the ones encountered on bulk Au. In our context, such normal metal-superconductor heterostructures are interesting not only for the ability to preserve graphene electronic properties, but also for the capability of hosting surface-assisted reactions.

Inspired by the seminal work of Tomanic {\it et al.},\cite{Tomanic2016} our work targets the Nb(110) substrate covered with a thin Ag buffer layer grown in UHV as a reliable superconducting platform.\cite{Tomanic2016} Using scanning tunneling microscopy (STM) and atomic force microscopy (AFM) at 4.7 K, we investigate Ullmann polymerization of 10,10'-dibromo-9,9'-bianthracene (DBBA) precursors. We characterize the synthesized nanostructures by AFM with CO-terminated tips and confirm proximity-induced superconductivity. We believe our results open a new route towards the study of topological superconductivity in atomically-precise NGs.

\section{Results and Discussion}

\begin{figure}[H]
\centering
 	\includegraphics[width=1.0\textwidth]{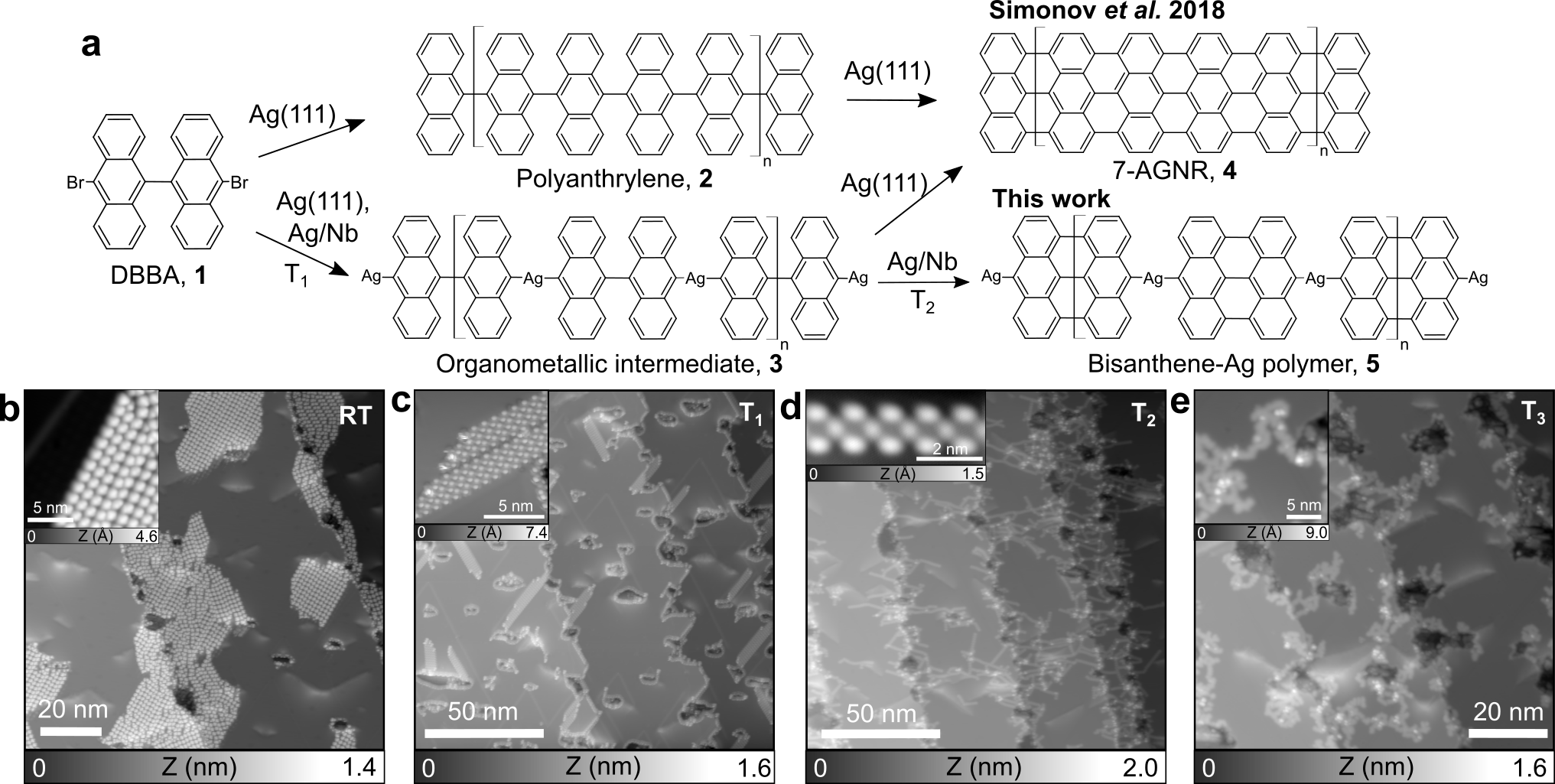}
 	\captionsetup{width=1.0\linewidth}
 	\caption{{\bf Bottom-up synthesis of NGs on the superconducting Ag/Nb(110) substrate.} 
{\bf a,} Hierarchical Ullmann polymerization leading to bisanthene-Ag chains as compared to literature. 
{\bf b-e,} Series of STM images showing the evolution of surface morphology as a function of substrate temperature. 
{\bf b,} Sublimating DBBA molecules ({\bf 1}) on Ag/Nb(110) leads to extended two-dimensional self-assemblies ($I_{\rm T}$= 1 pA, $V$ = 1.8 V. Inset: $I_{\rm T}$= 1 pA, $V$= -1.5 V). 
{\bf c,} Upon annealing to $T_{1}$= 75 $^{\circ}$C, molecular domains evolve to one-dimensional stacks of compounds {\bf 3} ($I_{\rm T}$ = 1 pA, $V$= -1.5 V. Inset: $I_{\rm T}$= 1 pA, $V$= 900 mV). 
{\bf d,} Annealing to $T_{2}$= 300 $^{\circ}$C leads to bisanthene-Ag chains {\bf 5} ($I_{\rm T}$= 1 pA, $V$= 1.8 V. Inset: $I_{\rm T}$= 1 pA, $V$= 900 mV). 
{\bf e,} The final thermal treatment to $T_{3}$= 390 $^{\circ}$C results in small NG domains ($I_{\rm T}$ = 1 pA, $V$  = 1.9 V. Inset: $I_{\rm T}$= 1 pA, $V$= 2 V).}
    \label{fig1}
\end{figure}

Our aim is to reproduce Ullmann polymerization of DBBA precursors (compound {\bf 1} in Fig.~\ref{fig1}a), which leads to 7-carbon-wide armchair GNRs (7-AGNRs) on Ag(111),\cite{Simonov2018,Jacobse2019} on the superconducting Ag/Nb substrate. To achieve this, we first investigate by STM the growth of Ag thin films on Nb(110) with thicknesses ranging from 0.2 to 5 monolayers (ML) (see Methods and Supplementary Information Figs. S2 and S3). Precursor {\bf 1} is then sublimated on the Ag/Nb substrate kept at room temperature, resulting in 2D self-assembled domains located mostly at step edges and defects (Fig. \ref{fig1}b). Upon annealing to $T_{1}$ = 75~$^{\circ}$C, dehalogenation of {\bf 1} is initiated, forming bundles of chains (Fig. \ref{fig1}c). Increasing the sample temperature to $T_{2}$ = 300~$^{\circ}$C then results in single polymeric chains (Fig.~\ref{fig1}d) while for $T_{3}$ = 390~$^{\circ}$C, molecular structures appear more curved and fused. The close-up STM image of each sub-product is shown in the insets of Figs.~\ref{fig1}b-e. We found the formation of these sub-products are independent of the Ag thickness explored so far.

\begin{figure}[t!]
\centering
 	\includegraphics[width=1.0\textwidth]{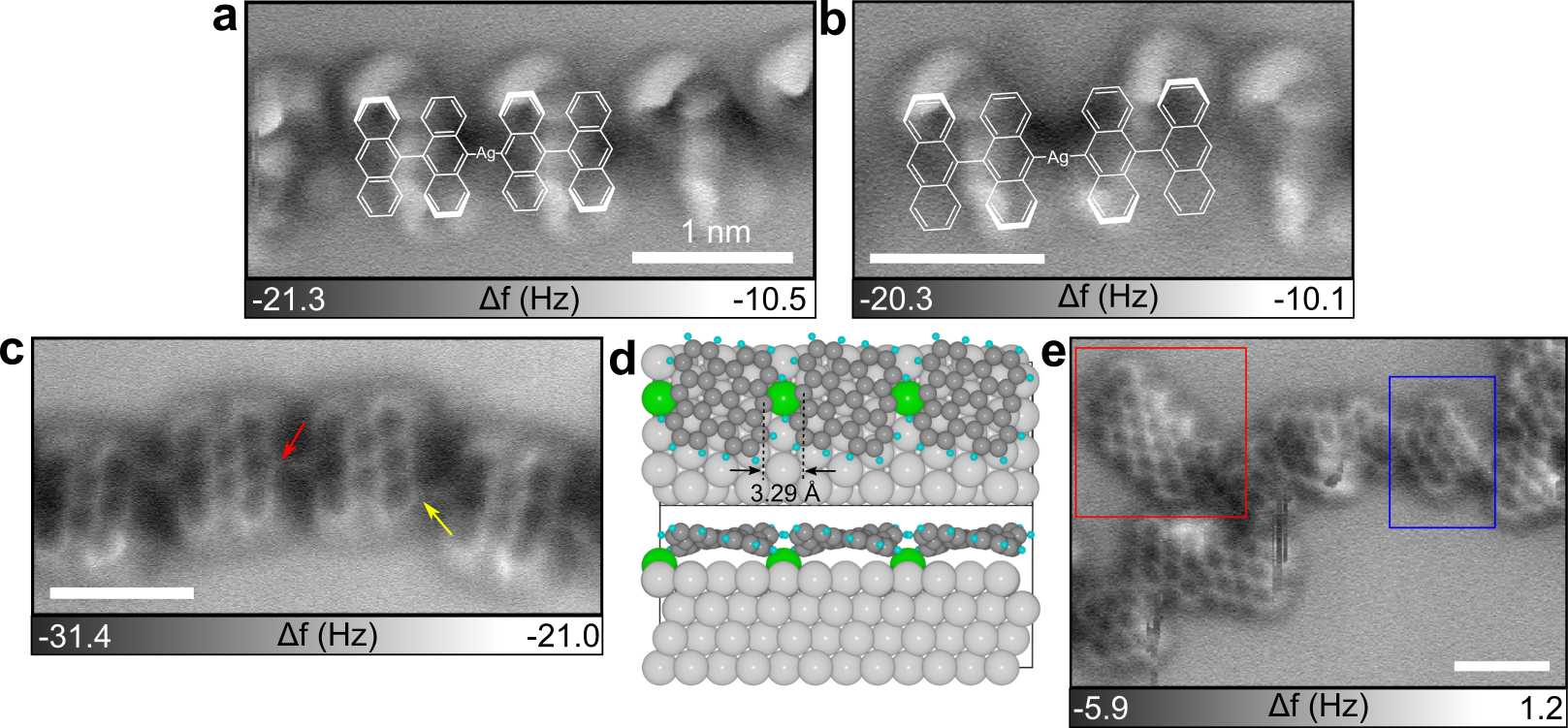}
 	\captionsetup{width=1.0\linewidth}
 	\caption{{\bf Structural characterization of reaction sub-products.} 
 	{\bf a,} AFM image of OM intermediates {\bf 3} showing a zigzag pattern. 
 	{\bf b,} AFM image of the OM intermediates {\bf 3} which have an armchair pattern fingerprint in AFM. Bold bonds in the corresponding K{\'e}kul{\'e} structures of {\bf a} and {\bf b} represent the most protruding part of the chain. Two buckling patterns of {\bf 3} might result from the Ag lattice-modulated effect. 
 	{\bf c,} AFM image of a bisanthene-Ag chain {\bf 5} where each monomer is linked by C-Ag-C bond. 
 	{\bf d,} Relaxed structure of the OM intermediate on Ag(111) optimized by DFT. The DFT simulation result shows the distance of 3.29~\AA~between bisanthene moieties. 
 	{\bf e,} AFM image of the irregularly fused bisanthenes (blue) leading to small 7-GNR segments (red) as well as irregular structures. All scale bars are 1 nm.}
    \label{fig2}
\end{figure}

To better understand the reaction steps, we elucidate the chemical structure of each sub-product using AFM imaging with CO-functionalized tips.~\cite{Gross2009} Figures~\ref{fig2}a and b display AFM images of zigzag and armchair chains respectively (see Figs. S4b and c for corresponding STM images). Both chain configurations are found as intermediates towards 7-AGNRs on noble metal surfaces.~\cite{Cai2010,Simonov2018} In our study, we assign both buckled structures to the formation of organometallic (OM) intermediates {\bf 3}, which is composed of bianthracene radicals and Ag surface adatoms. Due to the steric hindrance between anthracene moieties, only the topmost phenyl rings can be resolved by AFM (bold lines in the superimposed K{\'e}kul{\'e} structures of Figs.~\ref{fig2}a and b). Our assignment is different from the previous study on Ag(111) using the same precursor,~\cite{Simonov2018} in which the zigzag pattern was confirmed as polyanthrylene {\bf 2} after successful dehalogenation of {\bf 1} and formation of C-C bonds. We attribute the two different buckling patterns of {\bf 3} on Ag/Nb(110) to the accommodation to the distorted lattice of thin (2.5 ML) Ag films.

To enforce cyclodehydrogenation towards GNRs, we further anneal the sample to $T_{2}$ = 300~$^{\circ}$C. Exemplary STM/AFM images of the resulting product are shown in the inset of Fig.~\ref{fig1}d and Fig.~\ref{fig2}c. Surprisingly, cyclodehydrogenation only occurs within each bianthracene moiety, but not between adjacent bisanthene monomers (Fig.~\ref{fig2}c). By extracting the distance between adjacent bisanthenes, we find the interlinking bond length about 2.51${\pm}$0.07~\AA, which is too long to be a single C-C bond. Moreover, bright protrusions located at the interlinking position in the STM image (inset of Fig.~\ref{fig1}d) indicate conducting species conjugated in between. Based on these observations, we assign the synthesized structure of Fig.~\ref{fig2}c to the bisanthene-Ag polymer {\bf 5} even though Ag atoms are not clearly resolved by AFM.~\cite{Schulz2017} According to the relaxed structure of {\bf 5} on Ag(111) optimized by density function theory (DFT) (Fig.~\ref{fig2}d), surface Ag atoms are pulled out, yielding the distance of 3.29~\AA~between middle peripheral carbons of adjacent bisanthene monomers. This distance is still much larger than our measured bond length (2.51~\AA), which might explained by different lattice configurations between thin Ag films and Ag(111). Besides, shorter bond length could be caused by intrinsic limits of the AFM technique, such as the tilted CO molecule at the tip apex\cite{Gross2012} and slight drift during the slow scan.

With the confirmation of {\bf 5}, we stress that intermediate {\bf 2} is unlikely to be synthesized at this step of the reaction, since breaking a strong C-C bond between anthracene moieties and forming a much weaker OM (C-Ag) bond is not energetically favorable. Furthermore, bisanthenes are interlinked not only from the middle of the bisanthene edge (red arrow in Fig.~\ref{fig2}c) but also from the peripheral carbons (yellow arrow). This observation allows us to conclude that surface Ag adatoms on Ag/Nb(110) involve the substrate into the reaction, causing Ullmann-type reaction competes with surface-assisted dehydrogenative coupling (non-Ullmann) during the course of the reaction. Noteworthy is the synthesis of {\bf 5} from DBBA precursors on Ag/Nb(110), which is novel comparing to previous studies on Ag(111), suggesting that these two substrates do not share the identical catalytic reactivity.~\cite{Simonov2018,Jacobse2019}


Annealing the sample at $T_{3}$ = 390 $^{\circ}$C further promotes cyclodehydrogenation between adjacent bisanthenes. Instead of straight GNRs, we observe rather dendritic structures (Fig. \ref{fig1}e) which consist of irregularly fused NGs according to AFM imaging (Fig. \ref{fig2}e). Similar fused morphologies were also synthesized by Jacobse \textit{et al.}\cite{Jacobse2016} using chlorinated analogue precursors on Au(111). In their study, cyclodehydrogenation takes place within the molecule prior to dehalogenation, which leads to bisanthene radicals. Due to the steric hindrance between hydrogen atoms, bisanthene radicals bond with each other in an irregular manner. In our case, bisanthenes are metal-coordinated through three lateral positions (Fig. 2c), an alternative cyclodehydrogenation process is thus promoted which leads to fused nanographenes. Note also that annealing temperature\cite{Simonov2018} or heating rate\cite{Jacobse2019} could enable a better control of the reaction pathway towards a conventional Ullmann coupling. Despite the large involvement of the substrate, short segments of 7-AGNRs can be occasionally found as marked by the red rectangle in Fig. \ref{fig2}e. This thermodynamical aspect of the reaction will be explored in future works.

\begin{figure}[t!]
\centering
 	\includegraphics[width=1.0\textwidth]{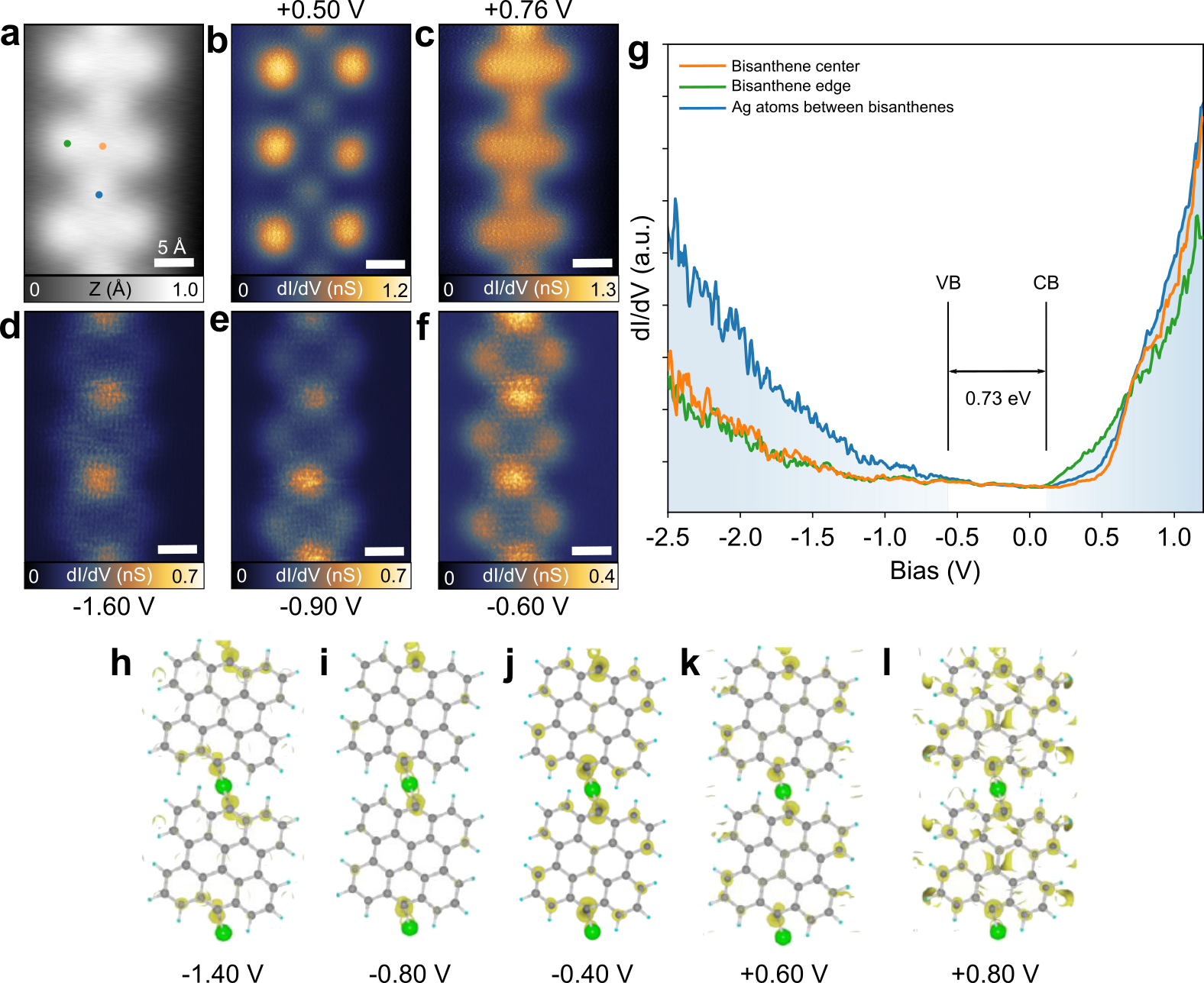}
 	\captionsetup{width=1.0\linewidth}
 	\caption{{\bf Experimental and simulated electronic properties of bisanthene-Ag chains \textbf{5}.} 
 	{\bf a,} STM image of a 3-unit bisanthene-Ag chain. Green, orange, and blue dots refer to the positions of $dI/dV$ spectra shown in {\bf g} ($I_{\rm t}$= 1 pA, $V$= 900 mV). 
 	{\bf b-f,} Series of $dI/dV$ maps at indicated bias. All scale bars are 5~\AA. 
 	{\bf g,} $dI/dV$ spectra measured at three representative positions of the bisanthene chain. Shaded areas refers to the onsets of CB and VB ($V$= 900 mV, $A_{\rm mod}$= 20 mV). 
 	{\bf h-l,} Simulated DOS of {\bf 5} at different energy levels shows consistence with the experimental results in {\bf b-f}.}
    \label{fig3}
\end{figure}

We next perform scanning tunneling spectroscopy (STS) on \textbf{5} and fused NGs to gain in-depth understanding about their electronic properties in combination with DFT calculations (see Methods). A series of differential conductance ($dI/dV$) spectra and $dI/dV$ mapping were acquired along the central chain \textbf{5} (Supplementary Information Fig. S5b) and across a bisanthene monomer (Supplementary Information Fig. S5c). Three representative $dI/dV$ spectra acquired at positions marked in the STM image of Fig.~\ref{fig3}a are shown in Fig. \ref{fig3}g. Along the chain axis, a shoulder at +0.76 eV is found on bisanthene monomers (orange) and Ag atomic sites (blue) which is absent at the bisanthene armchair edge (green). The latter has a resonance at +0.13 eV attributed to the conduction band (CB) onset. The frontier resonance of the valence band (VB) onset is assigned to -0.60 eV, allowing to extract an energy gap of about 0.73 eV. In comparison, DFT calculations of the OM structure in gas phase reveal a gap of 1.057 eV in relative agreement with experimental data. Note that  the band gap value extracted from STS measurements is typically reduced by an additional electron screening from the underlying metallic surface with respect to the band gap of the gas-phase polymer obtained by DFT.~\cite{Neaton2006}

The $dI/dV$ map acquired at +0.50 eV (Fig.~\ref{fig3}b) shows an increased density of states (DOS) over bisanthene edges and at Ag atomic sites while centers of bisanthene units are extinguished. The $dI/dV$ map above the VB edge (Fig.~\ref{fig3}d) shows maxima at Ag atoms and lateral termini of the bisanthene moiety, while the $dI/dV$ map acquired at +0.76 eV (Fig.~\ref{fig3}c) shows a continuous DOS over the entire chain. Despite the fact that DFT calculations cannot predict correctly the magnitude of the intrinsic band-gap of the polymeric chain, we find that the measured DOS of {\bf 5} on Ag/Nb has a similar trend with the calculated frontier orbitals of {\bf 5} on Ag(111) for the VB and CB band edges (see Figs.~\ref{fig3}h-l and Supplementary Information Fig. S6a), validating the character of the frontier orbitals predicted by DFT.

\begin{figure}[h!]
\centering
 	\includegraphics[width=0.75\textwidth]{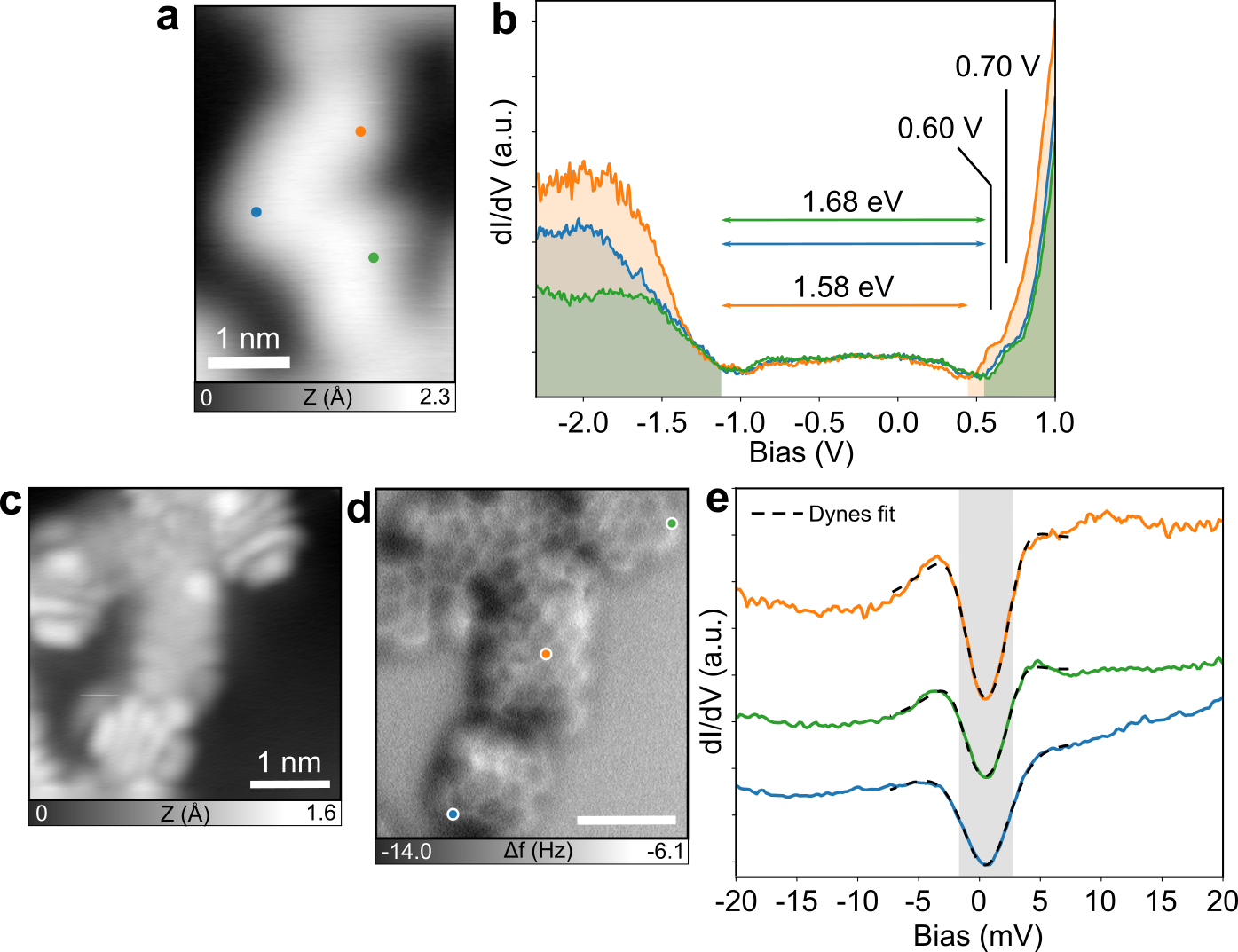}
 	\captionsetup{width=1.0\linewidth}
 	\caption{{\bf Proximity-induced superconductivity in fused nanographene.} 
 	{\bf a,} STM image of fused nanographene. Green, orange and blue dots refer to positions of $dI/dV$ spectra shown in {\bf b} ($I_{\rm t}$= 1 pA, $V$= 1.8 V). 
 	{\bf b,} $dI/dV$ spectra showing VB and CB onsets ($V$= 1.8 V, $A_{\rm mod}$= 20 mV). 
 	{\bf c,} STM image with a CO-functionalized tip at another nanographene area ($I_{\rm t}$= 1 pA, $V$= 150 mV). 
 	{\bf d,} Corresponding AFM image. The dots mark the positions of $dI/dV$ spectra shown in {\bf e}. 
 	{\bf e,} The estimated width of superconducting gaps are highlighted by the shaded area, indicating a robust proximity-induced superconductivity on the nanographene. Curves are vertically shifted for clarity ($V$= 150 mV, $A_{\rm mod}$= 2 mV).}
    \label{fig4}
\end{figure}

We last discuss the electronic properties of irregularly fused NGs (Fig. \ref{fig4}b). $dI/dV$ spectra acquired at different positions (Fig. \ref{fig4}a) share almost identical lineshape, allowing a gap estimation of 1.58-1.68 eV. We also carry out STS at 4.7 K with a metallic tip of the superconducting gap within the NG structure to compare local variations of the proximitized superconductivity (Fig. \ref{fig4}e). Systematic gaps were observed at various positions (Fig. \ref{fig4}d), which indicate the robustness of the proximity-induced superconductivity. Based on this observation, we also expect to observe the interaction with localized spins at zigzag edges, which may give rise to Yu-Shiba-Rusinov (YSR) states as pairwise peaks within the superconducting gap.~\cite{Rio2021} Unfortunately, we do not observe any clear signatures of YSR states at zigzag edges (blue and green dots in Fig. \ref{fig4}d). We first stress that the STS resolution in our experiments is smeared out by thermal broadening (4.7 K), and the use of Ag-coated metallic tip and the large lock-in modulation ($\approx$ 1 mV) are also likely to hamper the spectroscopic signature of YSR states. Besides, the absence of YSR states might also result from the passivation of unpaired spins by hydrogen contaminants or Ag adatoms. Future experiments at lower temperature are planned to address this issue in more details.

\section{Conclusion}
In conclusion, we fabricated a metal-superconductor heterostructure consisting of a Nb(110) substrate covered by thin Ag films, and find robust proximity-induced superconductivity on the Ag layer.~\cite{Tomanic2016} In contrast to the reactive Nb surface, we demonstrated by STM/AFM at low temperature that the Ag buffer layer is compatible with thermal-triggered on-surface reactions, including surface diffusion of molecules, dehalogenation, formation of C-C intra-monomer bonds as well as cyclodehydrogenation. The presence of surface Ag adatoms however alters the reaction pathway as compared to pristine Ag(111) and can lead to unexpected NG structures such as bisanthene-Ag polymeric chains or fused nanographene. Nevertheless, our results demonstrate an exciting starting point towards the general exploration of exotic electronic states in extended atomically precise NGs or metal-organic frameworks proximitized to a $s$-wave superconductor. This may open new routes towards the emergence of topological superconductivity in carbon-based nanostructures.

\section{Methods}
\paragraph*{Ag/Nb(110) preparation.}
Ag/Nb(110) substrates were prepared under UHV ($\approx$ 10$^{-10}$ mbar) following the protocol described here~\cite{Tomanic2016}. A Nb(110) substrate purchased from MaTeck GmbH was cleaned by cycles of Ar$^{+}$ sputtering and annealing using a home-made radio frequency (RF) heater. For sample annealing, we carried out five cycles of annealing up to $T$ $\geq$ 1600 $^{\circ}$C for 30 seconds followed by one minute of cooling. After careful degassing of the sample, this procedure allowed us to keep the pressure below 5$\times$10$^{-8}$ mbars during preparation. Ag films were grown on the Nb(110) surface kept at room temperature $via$ an e-beam evaporator (EFM3-Focus GmbH). Typical evaporation was about 30 min for a measured flux of about 30-40 nA. After Ag deposition, the sample was annealed with the RF heater to 550 $^{\circ}$C for 15 minutes in order to obtain flat and extended silver monolayers (see Supporting Information Figs. S2c-e and S3a). Temperatures were measured using a infrared pyrometer (Dias Infrared systems GmbH) with an emissivity $\xi$ of 0.12. 

\paragraph*{Molecule deposition.}
We used DBBA~\cite{Lee2017a} to perform Ullmann polymerization. DBBA was deposited at 170 $^{\circ}$C with the substrate remaining at room temperature. The molecule flux was measured by quartz microbalance, which showed 1.38 Å.min$^{-1}$ at 170 $^{\circ}$C. After the deposition, the sample was annealed with the RF heater at different temperatures for 10 minutes.

\paragraph*{STM/AFM experiments.}
All the samples were characterized at 4.7 K under UHV with the low-temperature STM/AFM provided by Omicron GmbH. The microscope is equipped with a qPlus sensor,~\cite{Giessibl2019} which has a natural oscillation frequency around 23.7 kHz and spring constant 1800 N.m$^{-1}$. In order to enhance the AFM resolution, the tip was functionalized with a CO molecule. CO functionalizing was done by depositing CO molecules on the cold surface ($\leq$ 15 K) and then gently indenting the tip on top of a CO molecule. $dI/dV$ spectra and maps were recorded with the lock-in amplifier with the modulation amplitude indicated in the caption. The width and the position of the superconducting gap were estimated by fitting the spectra with the thermally broadened Dynes function.

\paragraph*{Density functional theory (DFT).}
DFT calculations were performed using the Quickstep module of CP2K~\cite{Kuehne2020} using the gradient-corrected Perdew-Burke-Ernzerhof (PBE) exchange-correlation functional.~\cite{Perdew1996} Electron-nuclear interactions were described using Goedecker-Teter-Hutter (GTH) pseudopotentials with 11, 4 and 1 valence electron for Ag, C and H respectively. A molecularly-optimized~\cite{Lippert1997} double-zeta plus polarization (DZVP) basis set was used together with an auxiliary plane-wave basis set with kinetic-energy cutoff of 500 Ry. Reciprocal space was sampled using the $\Gamma$ point only. Table S1 shows details of the on-surface models used in these calculations. The self-consistent (SCF) calculations was terminated at an energy threshold of $10^{-6}$ Ha and structures were optimized until forces converged below $4.5\cdot 10^{-4}$ Hartree/Bohr.

\section{Acknowledgement}
Financial support from the Swiss National Science Foundation (SNSF grant 200021\_204053) and the Swiss Nanoscience Institute (SNI) is gratefully acknowledged. We also thank the European Research Council (ERC) under the European Union’s Horizon 2020 research and innovation programme (ULTRADISS Grant Agreement No. 834402). This project is under the scope of the QUSTEC program, which has received funding from the European Union's Horizon 2020 research and innovation program under the Marie Skłodowska-Curie grant number 847471. DFT calculations were financially supported by the SNSF Professorship PP00P2\_187185 and performed on UBELIX (http://www.id.unibe.ch/hpc), the HPC cluster at the University of Bern.

\section{Author Contributions}
E.M., R.P. and W.W. designed the experiments.
P.Z., S.X.L. and S.D. synthesized the molecule.
J.-C.L. performed STM/AFM experiments and analyzed the data.
X.W. and U.A. performed the DFT calculations.  
J.-C.L. wrote the manuscript with the help of R.P.
All authors discussed the results and revised the manuscript.

\mciteErrorOnUnknownfalse
\bibliography{reference}
\end{document}